\documentclass[aps,prl,twocolumn,groupaddress]{revtex4-1}

\usepackage[dvips]{graphicx}
\usepackage{amsmath}
\usepackage{amsfonts}
\usepackage{amssymb}
\usepackage[latin1]{inputenc}
\usepackage{xcolor}

\definecolor{darkblue}{rgb}{0, 0, 0.8}
\usepackage[colorlinks=true, breaklinks=true, linkcolor=darkblue, citecolor=darkblue, urlcolor=darkblue]{hyperref}
\newcommand{\doilink}[2]{\href{http://dx.doi.org/#1}{#2}}

\newcommand{\beq}{\begin{equation}}
\newcommand{\eeq}{\end{equation}}

\begin{document}

\title{Coherent Scattering of Near-Resonant Light by a Dense, Microscopic Cloud of Cold Two-Level Atoms: Experiment versus Theory}

\author{Stephan Jennewein}
\affiliation{Laboratoire Charles Fabry, Institut d'Optique Graduate School, CNRS, Universit\'e Paris-Saclay, 91127 Palaiseau Cedex, France.}
\author{Ludovic Brossard}
\affiliation{Laboratoire Charles Fabry, Institut d'Optique Graduate School, CNRS, Universit\'e Paris-Saclay, 91127 Palaiseau Cedex, France.}
\author{Yvan R.P. Sortais}
\affiliation{Laboratoire Charles Fabry, Institut d'Optique Graduate School, CNRS, Universit\'e Paris-Saclay, 91127 Palaiseau Cedex, France.}
\author{Antoine Browaeys}
\affiliation{Laboratoire Charles Fabry, Institut d'Optique Graduate School, CNRS, Universit\'e Paris-Saclay, 91127 Palaiseau Cedex, France.}
\author{Patrick Cheinet}
\affiliation{Laboratoire Aim\'e Cotton, CNRS, Universit\'e Paris-Sud, ENS Paris-Saclay, Universit\'e Paris-Saclay, B\^atiment 505, 91405 Orsay, France}
\author{Jacques Robert}
\affiliation{Laboratoire Aim\'e Cotton, CNRS, Universit\'e Paris-Sud, ENS Paris-Saclay, Universit\'e Paris-Saclay, B\^atiment 505, 91405 Orsay, France}
\author{Pierre Pillet}
\affiliation{Laboratoire Aim\'e Cotton, CNRS, Universit\'e Paris-Sud, ENS Paris-Saclay, Universit\'e Paris-Saclay, B\^atiment 505, 91405 Orsay, France}

\date{\today}

\begin{abstract}
We measure the coherent scattering of low-intensity, near-resonant light by a cloud of laser-cooled 
two-level rubidium atoms with a size comparable to the wavelength of light. 
We isolate a two-level atomic structure by applying a $300$\,G magnetic field. 
We measure both the temporal and the steady-state coherent optical  response of the cloud 
for various detunings of the laser and
for atom numbers ranging from 5 to 100. 
We compare our results to a microscopic coupled-dipole model and to a multi-mode, paraxial Maxwell-Bloch model. 
In the low-intensity regime, both models are in excellent agreement, thus validating the Maxwell-Bloch model. 
Comparing to the data, the models are found in very good agreement for relatively low densities ($n/k^3\lesssim 0.1$), 
while significant deviations start to occur at higher density.
This disagreement indicates that light scattering in dense, cold atomic ensembles is still not quantitatively understood, 
even in pristine experimental conditions. 
\end{abstract}

\maketitle

The study of near-resonant light scattering in dense  atomic ensembles has
seen  a renewed  interest recently with the recognition that it is more
subtle than one may think and is relevant for many applications~\cite{Guerin2016a}.
For example, experiments on hot vapors~\cite{Keaveney2012a} and dense
cold gases~\cite{Pellegrino2014,Jennewein2016a,Corman2017} triggered a debate on
the role of correlations between light-induced dipoles due to recurrent
scattering~\cite{Javanainen2014,Javanainen2016}. %,Javanainen2017}.
Also, resonant dipole interactions were predicted to prevent the observation of
Anderson localization of light in random atomic ensembles~\cite{Skipetrov2014}.
On an applied side, the ultimate resolution of optical clocks may be limited by the interactions
between light-induced dipoles during the probing
by resonant light~\cite{Chang2004,Campbell2017}.
Similarly, the study of quantum degenerate gases requires the measurement of {\it in situ}
properties of dense samples, and the dipole interactions could bias their
extraction~\cite{Chomaz2012}.  Nonetheless,
far from being a drawback, dipole interactions could be an asset,
allowing for example to build atom-light interfaces with large coupling
strength~\cite{Bettles2015,Bettles2016,Shahmoon2017} or platforms to
study topological properties~\cite{Perczel2017}, synchronization of dipoles~\cite{Zhu2015},
or dissipative spin models (e.g.~\cite{Kramer2015}).
Unwanted shifts could  even be  suppressed
when tailoring the  spatial arrangement of atoms~\cite{Chang2004,Kramer2016}.

All these questions have been investigated mainly using a classical description
where each atomic dipole responds linearly to  the laser field
and the fields scattered by all the other dipoles
(see e.g.~\cite{Morice1995,Ruostekoski1997,Bienaime2010,Pellegrino2014,
Jennewein2016a,Jenkins2016b,Schilder2016,Bromley2016}). This coupled-dipole
approach was tested in dilute ($n/k^3\ll 1$,
with $n$ the spatial density and $k$ the wavevector of light)
cold ensembles of rubidium, using a simplified scalar description of the
interaction~\cite{Bender2010,Bienaime2010,Roof2016,Guerin2016b}, or with strontium,
which features a $J=0-J=1$ structure similar to a classical dipole~\cite{Bromley2016,Zhu2016}.
There the agreement with the theory was satisfactory.
However, the case of dense cold atomic vapors of alkali ($n/k^3\sim 0.1-1$),
relevant for the above-mentioned questions, is
still problematic as the shift and broadening of the line for increasing density observed in
Refs.~\cite{Pellegrino2014,Jennewein2016a,Corman2017}
could not be reproduced quantitatively by the coupled-dipole theory.
Among explanations suggested for the disagreement are the role of the complex
internal structure of the atoms, where complicated 
internal dynamics could take place~\cite{Mueller2001,Kiffner2005,Munro2017},
and the failure of the low-intensity hypothesis for the driving field : in a dense
gas, an atom may be saturated  by the intense field radiated by a nearby one.
Including the saturation requires a density matrix approach,
and numerical simulations in the high intensity regime is challenging
as the size of the Hilbert space grows exponentially with the number of
atoms~\cite{Ostermann2012,Kramer2015,Kramer2016,Jones2017,Manzoni2017}.

\begin{figure}
\includegraphics[width=1\linewidth]{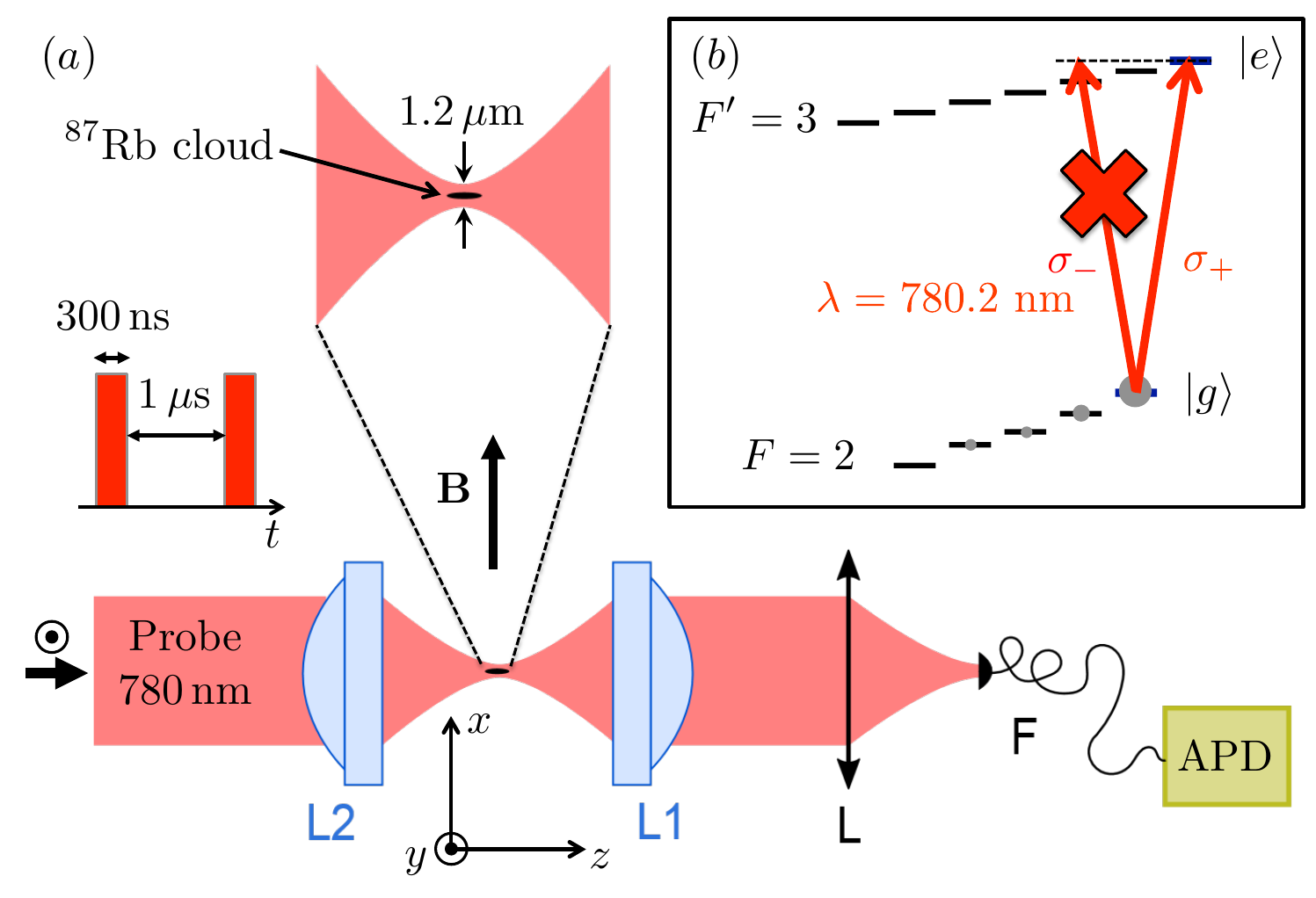}
\caption{(a) Experimental setup.
The atomic cloud is trapped at the focal point of two aspherical lenses L1 and L2 in a confocal configuration.
(L) lens focusing the light into a single-mode fiber (F) connected to an avalanche photodiode (APD).
The probe light is polarized along the $y$ direction.
(b) Level structure of $^{87}$Rb used in the experiment: the magnetic field $B=310$\,G
isolates a closed two-level transition between $|g\rangle$ and $|e\rangle$, detuned
from the closest transition by $12\Gamma$.
The $\sigma_-$ polarization component of the probe field thus does not
couple to the transition between $|g\rangle$ and $|e\rangle$.}
\label{Fig_setup}
\end{figure}

Here, we address experimentally and theoretically the two limitations raised above.
Firstly, we circumvent the problem of the atomic internal structure by producing a dense,
laser-cooled ensemble of {\it two-level} rubidium atoms. To do so, we isolate a two-level structure by
applying on the atoms a $\sim300$\,G magnetic field to lift the degeneracy of the Zeeman manifolds
(see Fig.~\ref{Fig_setup}). We measure the time-resolved {\it coherent} 
({\it i.e.} configuration-averaged) optical response
 of the cloud to a laser pulse nearly-resonant with that specific transition.
%The temporal response exhibits a faster dynamics than for a collection of independent atoms,
%indicating a super-radiant behavior.
Secondly, we investigate the low-intensity assumption
by developing a multi-mode Maxwell-Bloch model of the atom-field coupling
that can handle any intensity level, although neglecting correlations between dipoles. This model
agrees with the coupled-dipole theory,
and we validate the low-intensity hypothesis for the values used in the experiment.
Finally, comparing theory and experiment, we now find a very good agreement
for $n/k^3\lesssim 0.1$, thus validating the models.
For denser clouds, significant deviations appear, indicating that state-of-the-art models 
miss some  physical process in the description of light scattering in
dense and cold atomic ensembles, even in the pristine experimental conditions investigated here.

Our experimental setup relies on a cloud
of rubidium 87 atoms held in a microscopic dipole trap
(laser wavelength 940\,nm, $1/e^2$ waist 1.2\,$\mu$m, depth $1$\,mK),
with controllable average atom numbers  ranging between $\sim 1- 120$~\cite{Jennewein2016a}.
The temperature of the cloud is $150\,\mu$K, resulting in root-mean-square (rms) sizes of
the thermal distribution  $\sigma_r=0.3\lambda=230$\,nm and $\sigma_z=1.7\lambda=1.3\,\mu$m
in the radial and longitudinal directions
($\lambda=2\pi/\omega_0=780.2$\,nm is the wavelength of the D2 line of $^{87}$Rb
with linewidth $\Gamma=2\pi \times 6$\,MHz).
The atoms are initially prepared in a statistical mixture of
Zeeman states $M=0,\pm 1$ of the $F=1$ hyperfine ground-state manifold.
We polarize the sample in $|g\rangle=|5s_{1/2},F=2,M=2\rangle$ by sending a
combination of repumping and pumping light, both $\sigma_+$-polarized with
respect to the quantization axis set by a magnetic field,
and tuned respectively to the $(5s_{1/2},F=1)$ to $(5p_{3/2},F'=2)$
and $(5s_{1/2},F=2)$ to $(5p_{3/2},F'=2)$ transitions.
During this $1$\,ms pumping period, the magnetic field is set to $\sim 5$\,G.
We then lift the degeneracy of the Zeeman structures both in the $5s_{1/2}$ and $5p_{3/2}$
manifolds by increasing within $10$\,ms the magnetic field to $310$\,G.
This value is chosen to ensure that atoms in states other than $|g\rangle$ are
spectators with respect to the driving field and that the light scattered by atoms in
state $|g\rangle$ does not drive any transition other than
the one between $|g\rangle$ and $|e\rangle=|5p_{3/2},F'=3,M'=3\rangle$\,\cite{SM}.
We have measured that $80\%$ of the atoms are in state $|g\rangle$ at the end of this
polarization procedure, while the temperature remains unaffected. As a consequence, our largest
two-level atom peak density is $n=N/[(2\pi)^{3\over 2}\sigma_r^2\sigma_z]\approx 10^{14}$\,cm$^{-3}$,
corresponding to $n/k^3\approx0.15$. Here $N$ is the number of atoms in $|g\rangle$.

To probe the coherent optical response of the cloud, we use a laser beam (frequency $\omega$)
focused by a second aspherical lens (L2 in Fig.\,\ref{Fig_setup}a) in a
confocal configuration with respect to the one used to focus the dipole trap beam (L1)~\cite{Jennewein2016a}.
The probe at the position of the atomic cloud has a $1/e^2$ beam radius $1.20\pm0.05$\,$\mu$m.
It is linearly polarized along the $y$ axis, \textit{i.e.} perpendicular to the magnetic field.
In this way only the $\sigma_+$-component
of the probe field drives the transition from $|g\rangle$ to $|e\rangle$.
The probe intensity is kept low ($I/I_{\rm sat}\approx 0.02$, with $I_{\rm sat}=1.65$\,mW.cm$^{-2}$).
The transmitted part of the probe field,
sum of the laser field and of the field scattered by the cloud, is coupled to a single-mode
fiber connected to an avalanche photodiode (APD) operating in the single-photon counting mode.
We therefore measure the overlap of the transmitted field ${\bf E}({\bf r},t)$
with the mode of the fiber, assumed to be proportional to the laser
field ${\bf E}_{\rm L}({\bf r})$.
As the field in the forward direction is dominated by its {\it coherent}
part at low intensity~\cite{Footnote_forwardScatt},
we access the modulus square of the  coherent transfer function
$S(t,\omega) = \int \langle{\bf E}({\bf r},t,\omega)\rangle\cdot {\bf E}_{\rm L}^*({\bf r})dA / \int  |{\bf E}_{\rm L}({\bf r})|^2 dA$,
with $\langle\cdot\rangle$ denoting a configuration average.
The integral is performed over the area of the lens (L1).
This detection scheme is time resolved and allows studying the
dynamics of the scattering.
During the probing, the probe light is periodically switched on (for $300$\,ns) and off 1000 times,
and  interleaved with off and on periods (duration 1\,$\mu$s each) of recapturing of the cloud in the trap.
This recycling of the same cloud of atoms (although with different spatial positions)
keeps  heating and atom losses to less than $5\%$.
The $1000$-pulse sequence is repeated $200$ times, each
time with a new sample.

\begin{figure}
\includegraphics[width=0.95\linewidth]{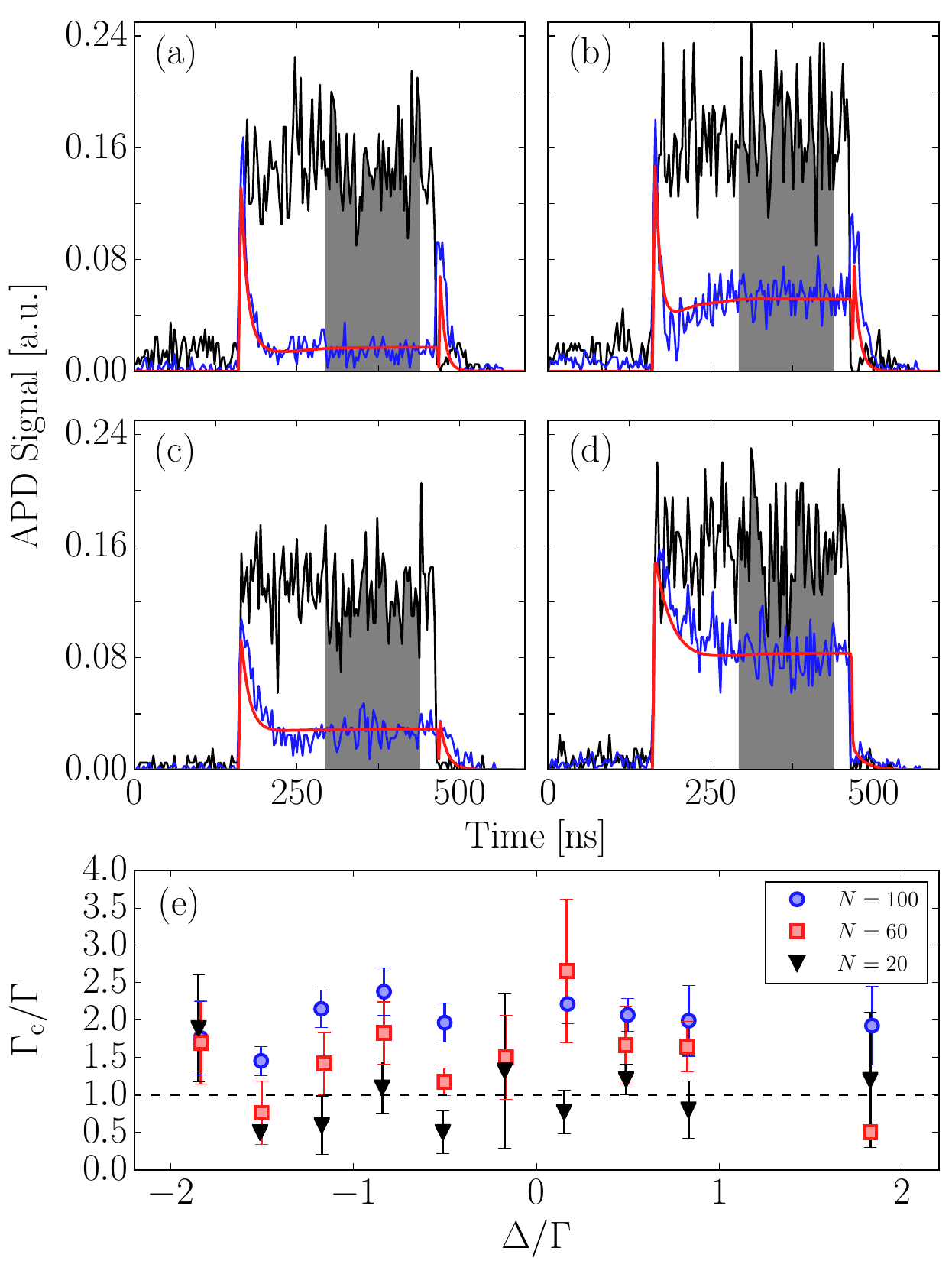}
\caption{(a-d) Examples of temporal responses of the cloud during and after illumination as
measured by the APD, for various detunings and atom numbers.
Black line: signal detected in the absence of atoms.
Blue line: response in the presence of atoms.
Red line: prediction of the time-dependent Maxwell-Bloch model (see text).
(a) $(N=100, \Delta=-0.14\Gamma)$,
(b) $(N=100, \Delta=-1.15\Gamma)$,
(c) $(N=60, \Delta=-0.5\Gamma)$,
(d) $(N=20, \Delta=-0.12\Gamma)$. The time bin of the detection is $1.5$\,ns.
(e) Decay rate $\Gamma_{\rm c}$ deduced from the fit of the temporal responses using Eq.~(\ref{Eq:phenomenological_fit}), as a function of the detuning $\Delta$ for various atom numbers.
Error bars are from the fit.}
\label{Fig_temporal_response}
\end{figure}

Figure~\ref{Fig_temporal_response} shows typical temporal signals recorded at the APD
with and without atoms in the trap. In the presence of atoms,
we observe the build-up of the destructive interference between
the laser field and the coherent field of the cloud due to its progressive polarization.
The signal then reaches a plateau, which defines empirically the steady-state regime.
Finally, for some parameters, we observe an after-pulse of coherent light after switching off the probe laser.
This pulse, which we did not see in our previous work using unpolarized sample~\cite{Jennewein2016a},
has been observed before in dilute gases~\cite{Toyoda1997,Chalony2011,Bettles2018}: it corresponds to the propagation
of the field coherently scattered by the atoms.

\begin{figure}
\includegraphics[width=0.95\linewidth]{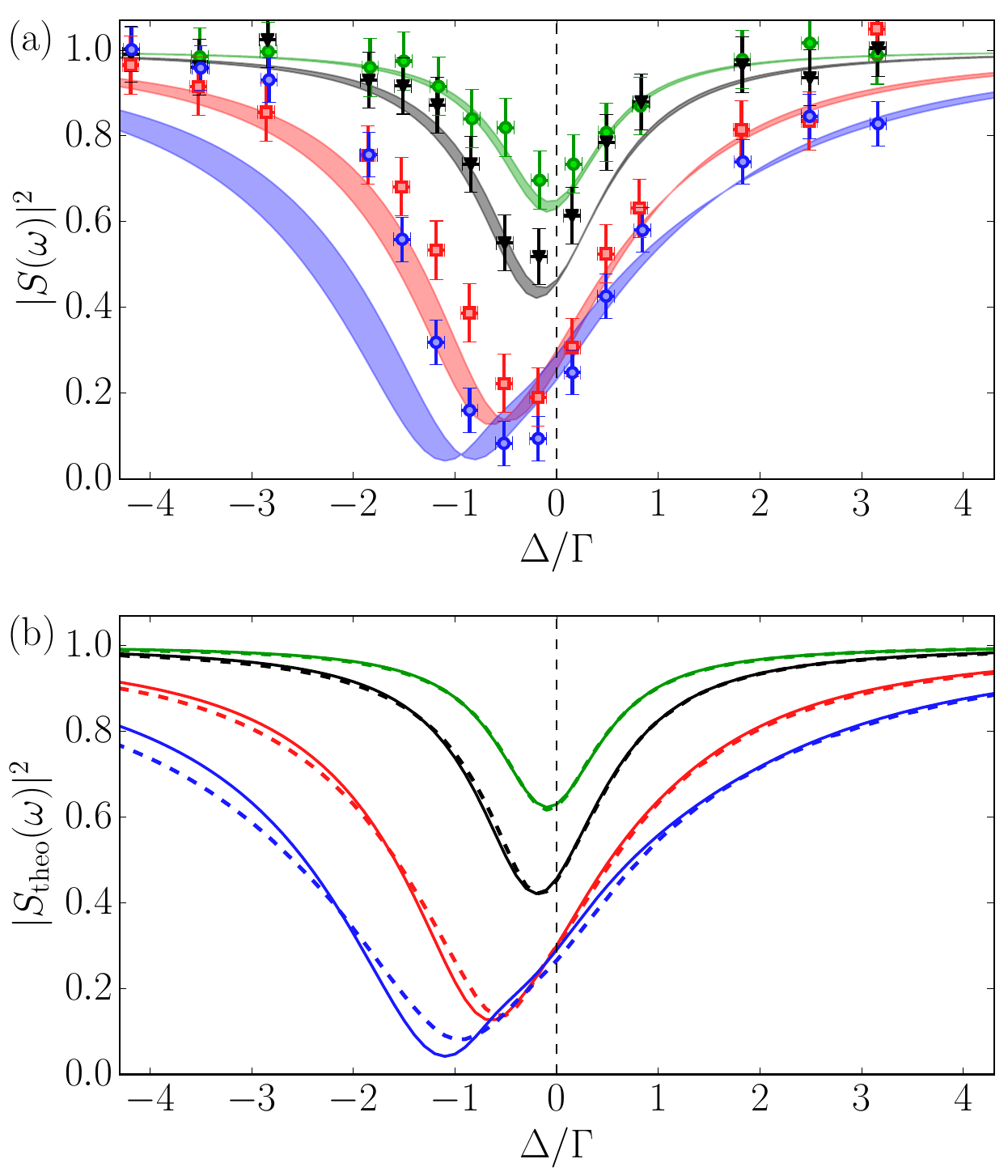}
\caption{(a) Transfer function $|{\cal S}(\omega)|^2$ as a function of the detuning
of the probe light for an ensemble of $N$ two-level atoms, for various atom numbers.
(Green, black, red, blue): $N=(10,20,60,100)$. Error bars are 2 s.e.m.
The thick solid lines are the
results of the multi-mode Maxwell-Bloch model (see text), {\it with no adjustable parameters}.
The thickness of the lines reflects the uncertainties on the cloud sizes and on the waist of the probe laser
(rms sizes $\sigma_r$ and $\sigma_z$ increased respectively by 10\% and 20\%, and
probe waist to 1.25~$\mu$m).
%The thickness of the line \yvan{illustrates the} uncertainty in the position $z_0$
%of the beam waist with respect to the cloud.
%The boundaries correspond to $z_0=0$ and $z_0=0.4z_{\rm R}$, \yvan{with $z_R=\pi w^2/\lambda\simeq 5.8\,\mu$m, the Rayleigh distance.}
(b) Comparison between the multi-mode Maxwell-Bloch model (dashed lines)
and the coupled-dipole model (solid lines) for the same parameters and atom numbers as in (a).}
\label{Fig_spectraS}
\end{figure}

Integrating the signal over $150$\,ns of the steady-state plateau (grey area in Fig.~\ref{Fig_temporal_response})
and normalizing to the signal detected without atoms yields the steady-state coherent transfer function.
We measured this function for various detunings $\Delta=\omega-\omega_0$ across
the resonance of the $|g\rangle$ to $|e\rangle$ transition and for $N$ ranging from $5$ to $100$
(see Fig.~\ref{Fig_spectraS}a). To compare the results to the multi-level
case of unpolarized samples~\cite{Jennewein2016a}, we fit the data by a Lorentzian profile (not shown).
This approach, although phenomenological, allows extracting a line shift, a line width and an amplitude.
Qualitatively, these parameters feature the same behavior
with the atom number as in the multi-level case, {\it i.e.} the shift and the broadening
increase linearly with $N$ and the amplitude saturates.
However the slope of the shift is about 2 times larger for the two-level
atom case and the saturation of the amplitude occurs at lower $N$.
This suggests that the internal atomic structure plays a role in the scattering.

We also fit the temporal response during the laser pulse by a phenomenological function
\beq\label{Eq:phenomenological_fit}
{\cal S}(t)= A|1 - {B\over 1+ 2i{\Delta-\Delta_{\rm c}\over \Gamma_{\rm c}}}
(1-e^{-i(\Delta-\Delta_{\rm c}) t-{\Gamma_{\rm c}\over 2}t})|^2,
\eeq
with $A$, $B$, $\Delta_{\rm c}$ and $\Gamma_{\rm c}$ as free parameters, assuming
that at a given detuning the laser excites only one eigenmode of the interacting dipoles,
with a shift $\Delta_{\rm c}$ and width $\Gamma_{\rm c}$.
The fit (not shown) is good for all detunings and atom numbers.
Figure~\ref{Fig_temporal_response}e
shows that when $N$ increases, $\Gamma_{\rm c}$ becomes larger than the radiative decay rate
of independent atoms by a factor $\sim 2$.
This indicates that the laser mainly couples to
super-radiant states involving a few atoms only~(see Fig.2 of ~\cite{Schilder2016}), as is also observed in dilute
cold atomic clouds~\cite{Roof2016,Guerin2016b}.

In order to model our data, we generalize the Maxwell-Bloch treatment of the propagation of a light field
in a medium consisting of $N$ two-level atoms\,\cite{GrossHaroche1983,Kiffner2005,Manzoni2017}
to the multi-mode case to account for the diffraction  by the microscopic cloud.
The derivation of the equations, starting from the master equation ruling the time-dependent
density operator $\sigma(t)$ of the atomic ensemble, are detailed in~\cite{SM}.
They govern the evolution of the atomic coherences and populations of each atom
in the presence of a driving field and of dipole-dipole interactions~\cite{Footnote_coupled_dipole}.
We then use a continuous medium approximation, {\it i.e.} we perform a configuration averaging:
we introduce the average coherence $\sigma_{ge}({\bf r},t)$, the average population $\sigma_{ee}({\bf r},t)$ and
the slowly-varying {\it coherent} field amplitude $\Omega_+({\bf r},t)= d\langle {\bf E}^*({\bf r},t)\cdot {\bf e}_+\rangle \exp{(ikz)}/\hbar$
propagating through the atomic sample at position ${\bf r}$
($d$ is the transition dipole and the unit vector ${\bf e}_+$ defines the $\sigma_+$-polarization~\cite{SM}).
We  account for the quasi one-dimensional shape of the cloud and demonstrate that $\Omega_+({\bf r},t)$ fulfills a paraxial equation. Finally, we include
the diffraction by decomposing the coherent field onto the Laguerre-Gauss basis:
$\Omega_+({\bf r},t)= \sqrt{\pi w^2\over 2}\sum_{q=0}^{\infty} \Omega_+^{(q)}(z,t){\rm LG}_q(r,z)$
($r,z$ are cylindrical coordinates)~\cite{SM}.
In the {\it low intensity regime}, we obtain a set of equations
coupling the slowly-varying, average coherence $\tilde{\sigma}_{ge}(z,t)=\sigma_{ge}(z,t)\exp(ikz)$
and the $\Omega_+^{(q)}(z,t)$ component of the coherent field:
\begin{align}
&{\partial \tilde{\sigma}_{ge}\over \partial t}=-\left({\Gamma\over 2}+i\Delta \right)\tilde{\sigma}_{ge}-i{\sqrt{\frac{\pi w^2}{2}} \over 2\pi \sigma_r^2}\sum_{p=0}^{\infty}f_p(z)\frac{\Omega_+^{(p)}}{2}\ ,\label{Eq:coupled_MBequations1}\\
&{\partial \Omega_+^{(q)}\over \partial z} +{1\over c}{\partial \Omega_+^{(q)}\over \partial t}=-i\sqrt{2\over \pi w^2}{3\pi\over 2k^2}\Gamma n \tilde{\sigma}_{ge} f^*_q(z)
\, e^{-{z^2\over 2\sigma_z^2}}\ ,\label{Eq:coupled_MBequations2}
\end{align}
with the initial conditions $\Omega_+(r,z=-\infty,t)=\Omega_{\rm L}(r,z=-\infty,t)$ (the laser field). The functions $f_q(z)$ are overlap integrals:
$f_q(z)= \int_0^\infty \exp\left( - {r^2\over 2\sigma_r^2}\right){\rm LG}_q(r,z)2\pi r dr$.
The steady-state response is solution of the equations:
\beq\label{Eq:MB_steady}
{d\Omega_+^{(q)}\over dz}=
-{3\Gamma n\over 4(k \sigma_r)^2}{f^*_q(z)\over \Gamma + 2i\Delta}\, e^{-{z^2\over 2\sigma_z^2}}\sum_{p=0}^{\infty}
f_p(z)\Omega_+^{(p)}\ ,
\eeq
where we have found that the sum  can be truncated after 10 modes for our experimental situation.
In this model, $|\Omega_+^{(0)}(z,t)|^2$ corresponds to the signal measured by the single-mode, fibered APD.
The equations for the {\it high intensity regime} are given in~\cite{SM}.
They rely on an  approximation neglecting
the correlations between the coherences and populations of different atoms.

We now compare the predictions of our model to a simulation of
the coupled-dipole equations~\cite{Bienaime2010,Bender2010,Pellegrino2014,Jennewein2016a,Bromley2016,Roof2016,Guerin2016b}
(see \cite{SM} for the equations in our geometry).
Figure~\ref{Fig_spectraS}b shows the results
of both models for  $N$ ranging from $5$ to $100$.
For  $N\le 60$ ($n/k^3\lesssim 0.1$),
they feature excellent agreement,
confirming the validity of our multi-mode Maxwell-Bloch model in the weak intensity limit.
For $N\ge 60$, residual deviations between the two models appear.
They are expected, as we approach the limits of validity of the continuous model
($n/k^3\lesssim0.2$)~\cite{Footnote_incoherent_scatt}.
Last, to investigate the validity of the low-intensity limit, we solve the
equations of the high intensity regime given in~\cite{SM} for the saturation parameter
$I/I_{\rm sat}=0.02$ used in the experiment. The results are indistinguishable from the solutions
of Eq.~(\ref{Eq:MB_steady}). Deviations between the predictions of the low and high-intensity regime
start to occur for $I/I_{\rm sat}\gtrsim 1$, as would be the case for non-interacting atoms.
This fact indicates that the saturation of an atom by the field radiated by a nearby one is not relevant.
Equivalently this means that sub-radiant modes, which may be saturated even at low intensity
since the associated saturation intensity scales with the decay rate, play a negligible role.

We finally compare our model to the measurements of the
steady-state and temporal coherent response of the cloud.
As the multi-mode Maxwell-Bloch approach agrees with
the coupled-dipole model for the experimental parameters, we are in fact comparing both
models to the data.
Figure~\ref{Fig_spectraS}a shows the results of the Maxwell-Bloch
model together with the data in steady state.
Contrarily to our previous work on unpolarized samples~\cite{Jennewein2016a}),
we observe here a  good agreement between the
data and the model {\it with no adjustable parameters} for $N\lesssim 60$, indicating that
light scattering is quantitatively understood for $n/k^3\lesssim 0.1$.
For denser clouds, the theory matches the data on the blue side of the resonance
but deviates significantly on the red side.
This discrepancy for $n/k^3\gtrsim 0.1$ suggests that extra physical effects must be included
to reproduce quantitatively the experiment.
To study the dynamics of the transfer function during
and after the pulse (see examples in Fig.~\ref{Fig_temporal_response})
we solve the time-dependent set of
equations~(\ref{Eq:coupled_MBequations1},\ref{Eq:coupled_MBequations2}).
Examples of results are shown as solid lines in
Fig.~\ref{Fig_temporal_response}, together with the data.
We observe that all features of the dynamics are reproduced by
the model, in particular the flash of light appearing after the probe laser
has been turned off~\cite{Footnote_timedependent}.
%This good agreement between time-dependent data and the multi-mode model
%is another check of its validity for our range of densities.

In conclusion, we have studied the coherent optical response of a dense cloud of cold,
two-level atoms excited in the low-intensity regime near an atomic resonance. 
We have compared the experiment with a model based on a multi-mode
Maxwell-Bloch approach. This model agrees with the coupled-dipole model, 
which validates the low-intensity assumption, and is able
to reproduce quantitatively the measured temporal and steady-state response for 
$n/k^3\lesssim 0.1$. For $n/k^3\gtrsim 0.1$, we observe discrepancies with
both models. Owing to the clean experimental situation with the isolation of two-level atomic 
structure, the only remaining extra effect is probably the residual motion of  the atoms,
perhaps enhanced by the forces between atoms. This finding calls for more theoretical studies
(e.g. using the formalism developed in \cite{Zhu2016}),
as it is key to the realization of the proposals mentioned in the introduction.

\begin{acknowledgments}
We thank J.-J.~Greffet, J.~Ruostekoski, T.~Gallagher, D.~Comparat and E.~Arimondo for discussions.
We acknowledge support by the ``Investissements d'Avenir'' LabEx PALM (project ECONOMIQUE)
by the R\'egion \^Ile-de-France in the framework of DIM Nano-K (project LISCOLEM), and
by the EU (Grant No. H2020 FET-PROACT Project RySQ).
\end{acknowledgments}


\begin{thebibliography}{80}
\bibitem{Footnote_coupled_dipole}
In the limit of a weak driving field, the equations on
the $N$ coherences $\sigma_{ge}^j$ of each atom are equivalent to the coupled-dipole equations.
\bibitem{Footnote_incoherent_scatt}
We have checked by solving the coupled-dipole equations that the
incoherent scattering (not included in the Maxwell-Bloch model) is negligible, and thus cannot explain the discrepancies between the two
models for $N\geq 60$.
\bibitem{Footnote_timedependent}
In the simulation
the only free parameters are the overall amplitude and the detuning, which
is chosen such that the theoretical steady-state transmission matches the measured one.
\end{thebibliography}
\end{document}

\title{Supplemental Material:  Coherent Scattering of Near-Resonant Light by a Dense, Microscopic Cloud of Cold Two-Level Atoms: Experiment versus Theory}

\author{Stephan Jennewein}
\affiliation{Laboratoire Charles Fabry, Institut d'Optique Graduate School, CNRS, Universit\'e Paris-Saclay, 91127 Palaiseau Cedex, France.}
\author{Ludovic Brossard}
\affiliation{Laboratoire Charles Fabry, Institut d'Optique Graduate School, CNRS, Universit\'e Paris-Saclay, 91127 Palaiseau Cedex, France.}
\author{Yvan R.P. Sortais}
\affiliation{Laboratoire Charles Fabry, Institut d'Optique Graduate School, CNRS, Universit\'e Paris-Saclay, 91127 Palaiseau Cedex, France.}
\author{Antoine Browaeys}
\affiliation{Laboratoire Charles Fabry, Institut d'Optique Graduate School, CNRS, Universit\'e Paris-Saclay, 91127 Palaiseau Cedex, France.}
\author{Patrick Cheinet}
\affiliation{Laboratoire Aim\'e Cotton, CNRS, Universit\'e Paris-Sud, ENS Paris-Saclay, Universit\'e Paris-Saclay, B\^atiment 505, 91405 Orsay, France}
\author{Jacques Robert}
\affiliation{Laboratoire Aim\'e Cotton, CNRS, Universit\'e Paris-Sud, ENS Paris-Saclay, Universit\'e Paris-Saclay, B\^atiment 505, 91405 Orsay, France}
\author{Pierre Pillet}
\affiliation{Laboratoire Aim\'e Cotton, CNRS, Universit\'e Paris-Sud, ENS Paris-Saclay, Universit\'e Paris-Saclay, B\^atiment 505, 91405 Orsay, France}

\date{\today}

\maketitle

This Supplemental Material contains four Sections. In the first one we give more details
about the choice of the magnetic field at which we perform the light scattering experiment.
In Section~\ref{Sec:details_MB_low_field}  we detail the multi-mode Maxwell Bloch model and
derive Eq.(1-3) of the main text. This part deals with the low intensity regime.
In Section~\ref{Sec:details_MB_strong_field}, we give the
expressions valid for any intensity of the driving field.
In the last Section, we derive the coupled dipoles
equations for our particular experimental configuration.

\section{Choice of the magnetic field for polarizing the atomic sample}\label{Sec:choice_B_field}

The value of the final magnetic field ($310$\,G) at which we perform the experiment
is governed by two considerations.
First, we want that the atoms remaining in Zeeman states $|5s_{1/2},F=2, M\neq2\rangle$ do not
interact with a probe laser tuned on the transition from $|g\rangle=|5s_{1/2},F=2, M=2\rangle$
to $|e\rangle=|5p_{3/2},F'=3,M'=3\rangle$.
This requires that the closest $\sigma_+$-transition frequency
is detuned by several linewiths $\Gamma$ ($\Gamma=2\pi \times 6$\,MHz for the D2 transition)
due to the Zeeman splitting. We have calculated the frequency
of the various optical transitions in our regime, intermediate between Zeeman and Paschen-Back,
and found that the nearest $\sigma_+$-transition is detuned by
$12\Gamma$ for a field of $310$\,G, thus making the atoms in states other than $|g\rangle$ spectators.
The second consideration comes from the dipole-dipole interaction itself:
we want that the field elastically scattered by any atoms of the cloud
at a given position (and thus with any possible polarization) does not drive
any transition other than the one between $|g\rangle$ and $|e\rangle$ (see Fig.~1(b) of the main text).
This implies that the magnetic splitting is larger than the dipole-dipole transition strength,
which is here as large as $3\Gamma$ (see Fig. 3 of the main text).
For $310$\,G the transition between $|g\rangle$ and $|5p_{3/2},F'=3,M'=2\rangle$
is detuned by $70\Gamma$ with respect to the transition between
$|g\rangle$ and $|e\rangle$, thus fulfilling the criterion.

\section{Derivation of the multi-mode Maxwell-Bloch model: low intensity case.}\label{Sec:details_MB_low_field}

In this Section, we  derive the multi-mode Maxwell-Bloch model that leads to Eqs.(1-3) of the main text.
In particular, we demonstrate that the coherent ({\it i.e.} configuration averaged) field $\Omega_+(r,z,t)$
is solution of a paraxial equation (Eq.\,\ref{Eq:Paraxial_eq}), a fact which is usually imposed without demonstration.
We consider here mainly the regime of low intensity of the driving field, leaving the high-intensity regime to Sec.~\ref{Sec:details_MB_strong_field}.

The model starts from the master equation ruling the density  operator $\sigma(t)$ describing a collection of $N$, two-level atoms with ground and excited states $|g\rangle$ and $|e\rangle$ respectively (transition frequency $\omega_0$)~\cite{GrossHaroche1983,Kiffner2005} in interaction with the modes of the vacuum field. In interaction representation
%and using only the Born approximation (the usual
%Markov approximation {\it is not} done here)
this equation reads, in the absence of driving field:
\begin{widetext}
\begin{align}\label{Eq:Born_without_Markov}
\frac{d\sigma(t)}{dt} &  =-\frac{\Gamma}{2}\sum_{j=1}^N
[r_j^+r_j^{-},\sigma(t)]_{+}-2r_j^-\sigma(t)r_{j}^{+} \nonumber\\
&  -\frac{3}{8}\Gamma\underset{l\neq j}{\sum}\left(  r_j^{+}r_l^{-}\sigma(t-\frac{R_{jl}}{c})-r_l^{-}\sigma(t-\frac{R_{jl}}{c})r_j^{+}\right)
\left(   -i{1+\cos^{2}\theta_{jl}\over kR_{jl}}
-\left(  1-3\cos^{2}\theta_{jl}\right) {kR_{jl}  +i \over (kR_{jl})^3}\right)
\exp\left( ikR_{jl}\right)  \nonumber\\
&  +\frac{3}{8}\Gamma\underset{l\neq j}{\sum}\left(r_j^{-}r_l^{+}\sigma(t-\frac{R_{jl}}{c})-r_l^{+}\sigma(t-\frac{R_{jl}}{c})r_j^{-}\right)
\left(  i {1+\cos^{2}\theta_{jl}\over kR_{jl}}  -(1-3\cos^{2}\theta_{jl})
\frac{kR_{jl}-i  }{(kR_{jl})^3}\right)\exp\left( -ik R_{jl}\right)\ .
\end{align}
\end{widetext}
Here, $[\cdot,\cdot]_+$ is the anti-commutator of two operators, $\Gamma$ is the decay rate of state $|e\rangle$, related to the dipole matrix element $d$ between $|e\rangle$ and $|g\rangle$ by $\Gamma= {k^3 d^2\over 3\pi\epsilon_0\hbar}$, with $k=\omega_0/c$. The atoms  $j$ and $l$ are located at positions ${\bf R}_j$ and ${\bf R}_l$ respectively, and the inter-particle distance is $R_{jl}= |{\bf R}_{jl}|$, with ${\bf R}_{jl}={\bf R}_l-{\bf R}_j$. Also  $\theta_{jl}$ is the angle between the vector ${\bf R}_{jl}$ and the quantization axis ${\bf e}_x$ (see geometry in Fig.~1(a) of the main text). The atomic dipoles are circularly polarized ($\sigma_+$) in the $yz$ plane. Finally, the raising and lowering operators $r_j^{\pm}$ for atom $j$ are defined as: $r_j^+=|e\rangle_j\langle g|_j$ and $r_j^-=|g\rangle_j\langle e|_j$. These equations are established by choosing the quantization axis as the direction of the applied magnetic field: the angular dependence would be  different with the quantization axis along the propagation axis.

When deriving Eq.\,(\ref{Eq:Born_without_Markov}), we have used the Born approximation, which means that only one spontaneous emission event occurs during the characteristic time of the evolution of the system $\sim 1/(N\Gamma)$. 
%However, we {\it did not} perform the Markov approximation as is usually done~\cite{GrossHaroche1983,Kiffner2005}. This is the reason for the presence of the retarded times $t- R_{jl}/ c$. 
However, we kept the retarded times $t- R_{jl}/ c$, {\it i.e.} we did not performed the Markov approximation.  
This leads to the  simple form of the Maxwell-Bloch equations (Eq.(1-3) of the main text), that would not be possible to obtain without keeping them. 
%To the best of our knowledge this fact has been overlooked so far, and the equations that we present here are original.

We obtain the  density operator for the atom $j$ by tracing over the $l\neq j$ other atoms, $\sigma^j(t)={\rm Tr}_{l\neq j}[\sigma(t)]$, hence:
\begin{widetext}
\begin{align}\label{Eq:density_operator}
\frac{d\sigma^j}{dt}(t)   & =-\frac{\Gamma}{2}\left(  [r_j^+r_j^-,\sigma^j(t)]_+ -2r_j^-\sigma^j(t)r_j^{+}\right) \nonumber\\
& +i\frac{3}{8}\Gamma\underset{l\neq j}{\sum}
\left[  \sigma_{eg}^l(t-\frac{R_{jl}}{c})r_j^+,\sigma^j(t-\frac{R_{jl}}{c})\right]
\left(  {1+\cos^{2}\theta_{jl}\over kR_{jl}}+
\left(  1-3\cos^{2}\theta_{jl}\right)
\frac{1-ikR_{jl}}{\left(kR_{jl}\right)  ^{3}}\right)\exp\left(  ikR_{jl}\right) \nonumber\\
& +i\frac{3}{8}\Gamma\underset{l\neq j}{\sum}
\left[ \sigma_{ge}^l(t-\frac{R_{jl}}{c})r_j^-,\sigma^j(t-\frac{R_{jl}}{c})\right]
\left(  {1+\cos^{2}\theta_{jl}\over kR_{jl}}+\left(  1-3\cos^{2}\theta_{jl}\right)
\frac{1+ikR_{jl}}{\left(kR_{jl}\right)  ^{3}}\right)\exp\left(  -ikR_{jl}\right)\ ,
\end{align}
\end{widetext}
with $[\cdot,\cdot]$ the commutator of two operators.

We now add the driving field of the laser ${\bf E}_{\rm L}({\bf r})=E_{\rm L}({\bf r}) {\bf e}_y$
propagating in the direction ${\bf e}_z$ with wavevector $k_{\rm L}=2\pi /\lambda$. In principle $k_{\rm L}$ should not be confused with $k=\omega_0/c$. However, as we operate
close to the atomic resonance, we will take in the calculations below $k=k_{\rm L}$. The laser beam having a Gaussian spatial profile,
\beq\label{Eq:laser_fieldE}
E_{\rm L}(r,z) =
E_0 \frac{1}{1+iz/z_{\rm R}}
\exp\left[{i\over 2}{kr^2\over z-iz_{\rm R}}\right]\exp(ikz)\ ,
\eeq
with $z_{\rm R} = k w^2/2$ the Rayleigh distance. $(r,z)$ are the radial and longitudinal cylindrical coordinates.

Also, we define the Rabi frequency associated to the slowly-varying envelope of the laser amplitude by $\Omega^*_{\rm L}~=~d \left({\bf E}_{\rm L}\cdot{\bf e}^*_+\right)\exp(-ikz)/\hbar$ :
\begin{equation}\label{Eq:laser_field}
\Omega_{\rm L}(r,z) =
\Omega_0 {iz_{\rm R}\over z+iz_{\rm R}}
\exp\left[-{i\over 2}{kr^2\over z+iz_{\rm R}}\right]\ ,
\end{equation}
with ${\bf e}_+=-({\bf e}_y+i{\bf e}_z)/\sqrt{2}$ and $\Omega_0=-dE_0/\sqrt{2}$. We obtain the equation relating the coherence $\sigma_{ge}^j(t)$ of atom $j$ and the ground and excited populations, $\sigma_{gg}^j(t)$ and
$\sigma_{ee}^j(t)$ respectively :
\begin{align}\label{Eq:coherence_i_Appendix}
&\frac{d\sigma_{ge}^j(t)}{dt}   =-\left(  \frac{\Gamma}{2}+i\Delta\right)\sigma_{ge}^j(t)
+i\frac{\Omega_{\rm L}(r_j,z_j)}{2}\exp\left(-ikz_j\right)\left(\sigma_{ee}^j(t-\frac{R_{jl}}{c})-\sigma_{gg}^j(t-\frac{R_{jl}}{c})\right)\nonumber\\
&+\frac{3}{8}i\Gamma\underset{l\neq j}{\sum
}\sigma_{ge}^l(t-\frac{R_{jl}}{c})
\left[ \sigma_{ee}^j(t-\frac{R_{jl}}{c})-\sigma_{gg}^j(t-\frac{R_{jl}}{c})\right]
\left[  {1+\cos^{2}\theta_{jl}\over kR_{jl}}+\left(  1-3\cos^{2}\theta_{jl}\right)
\frac{1+ikR_{jl}}{\left(kR_{jl}\right)  ^{3}}\right]\exp\left(  -ikR_{jl}\right)\ ,
\end{align}

Similarly, the equation on the excited state population is:
\begin{align}\label{Eq:population_i_Appendix}
{d\sigma_{ee}^j(t)\over dt} &  =-\Gamma\sigma_{ee}^j(t)
+\frac{3}{8}i\Gamma\underset{l\neq j}{\sum}\sigma_{eg}^l(t-\frac{R_{jl}}{c})\sigma_{ge}^j(t-\frac{R_{jl}}{c})\left[  {1+\cos^{2}\theta_{jl}\over kR_{jl}}
+\left(  1-3\cos^{2}\theta_{jl}\right)  \frac{1-ikR_{jl}}{\left(kR_{jl}\right)^{3}}\right]\exp\left(  ikR_{jl}\right)  \nonumber\\
&  -\frac{3}{8}i\Gamma\underset{l\neq j}{\sum}\sigma_{ge}^l(t-\frac{R_{jl}}{c}) \sigma_{eg}^j(t-\frac{R_{jl}}{c})
\left[  {1+\cos^{2}\theta_{jl}\over kR_{jl}}+
\left(  1-3\cos^{2}\theta_{jl}\right) \frac{1+ikR_{jl}}{\left(kR_{jl}\right)  ^{3}}\right] \exp\left(  -ikR_{jl}\right)\nonumber\\
&  +i\frac{\Omega^{\ast}_{\rm L}(r_j,z_j)}{2}
\exp\left(ikz_j\right) \sigma_{ge}^j(t)-i\frac{\Omega_{\rm L}(r_j,z_j)}{2}\exp\left( -ikz_j\right)\sigma_{eg}^j(t)\ .
\end{align}

As explained in the main text, we restrict ourselves to the weak-field limit, hence $\sigma_{ee}^j\approx 0$ and $\sigma_{gg}^j\approx1$. The model remains valid at high intensity by keeping the equation on the evolution of the population. We do not detail the calculations in the high intensity case, and will just give the results in Sec.~\ref{Sec:details_MB_strong_field}.  From now on, we consider the equation on the coherence in { \it the weak driving field limit}:
\beq\label{Eq:coherence_i_weakfield}
\frac{d\sigma_{ge}^j(t)}{dt}   =-i\frac{\Omega_{\rm L}(r_j,z_j)}{2}\exp\left(-ikz_j\right)-\left(\frac{\Gamma}{2}+i\Delta\right) \sigma_{ge}^j(t)-i\frac{3}{8}\Gamma\underset{l\neq j}{\sum}
\sigma_{ge}^l(t-\frac{R_{jl}}{c}) f(R_{jl},\theta_{jl})\exp\left(-ikR_{jl}\right)\,
\eeq
%{\color{red} La bonne equation qui redonne les dipoles couples avec $d_j=d\, \sigma_{ge}^j$
%\beq\label{Eq:coherence_i_weakfield}
%\frac{d\sigma_{ge}^j(t)}{dt}   =-i\frac{\Omega_{\rm L}(r_j,z_j)}{2 \sqrt{2}}
%\exp\left(ikz_j\right)
%-\left(  \frac{\Gamma}{2}-i\Delta\right)  \sigma_{ge}^j(t)
%+i\frac{3}{8}\Gamma\underset{l\neq j}{\sum}
%\sigma_{ge}^l(t-\frac{R_{jl}}{c}) f(R_{jl},\theta_{jl})\exp\left(  ikR_{jl}\right)\,
%\eeq}
where we have introduced the function:
\beq\label{Eq:f}
f(R,\theta) = \frac{1+\cos^2\theta}{kR}+\left(1-3\cos^2\theta\right) \frac{1+ikR}{\left(kR\right)^3}\ .
\eeq
%{\color{red} La bonne equation:
%\beq\label{Eq:f}
%f(R_{jl},\theta_{jl}) = {1+\cos^{2}\theta_{jl}\over kR_{jl}}+
%\left(  1-3\cos^{2}\theta_{jl}\right) \frac{1-ikR_{jl}}{\left(kR_{jl}\right)  ^{3}}\ .
%\eeq}
As the coherence $\sigma_{eg}^j(t)=[\sigma_{ge}^j(t)]^*$ is related to the complex amplitude of the average dipole of atom $j$ by $\langle d_j(t)\rangle~=~d\, \sigma_{eg}^j(t)$,
Eqs.~(\ref{Eq:coherence_i_weakfield}) in steady-state are identical to the coupled-dipole equations derived from classical electrodynamics (see Sec.~\ref{Appendix:detailed_coupled_dipoles}).

We now introduce the equations for the electromagnetic field. The total field at the location of atom $j$ is the superposition of the laser field and the field radiated by all the other atoms. Expressed in term of the slowly-varying Rabi frequency defined by $\Omega^* = d({\bf E}\cdot {\bf e}^*_+) \exp(-ikz)/\hbar$, the field driving the $\sigma_+$-polarized dipoles is therefore given by :
\begin{widetext}
\beq\label{Eq:field_total}
\Omega({\bf R}_j,t)= \Omega_{\rm L}(r_j,z_j)+{3\over8}\Gamma\exp\left(ikz_j\right)\underset{l\neq j}{\sum}\sigma_{ge}^l(t-\frac{R_{jl}}{c})\exp\left(-ikR_{jl}\right)f(R_{jl},\theta_{jl})\ .
\eeq
\end{widetext}
This expression is valid for any amplitude (weak or strong) of the laser field. Decomposing the field into a component $\Omega_+$ propagating in the direction ${\bf k}_{\rm L}$
\beq\label{Eq:Omega_+}
\Omega_+({\bf R}_j,t)= \Omega_{\rm L}(r_j,z_j)+{3\over4}\Gamma\exp\left(ikz_j\right)\underset{l< j}{\sum}\sigma_{ge}^l(t-\frac{R_{jl}}{c})\exp\left(-ikR_{jl}\right) f(R_{jl},\theta_{jl})\ ,
\eeq
and $\Omega_-$ propagating in the $-{\bf k}_{\rm L}$ direction
\beq\label{Eq:Omega_-}
\Omega_-({\bf R}_j,t)={3\over4}\Gamma\exp\left(ikz_j\right)\underset{l> j}{\sum}\sigma_{eg}^l(t-\frac{R_{jl}}{c})
\exp\left(-ikR_{jl}\right)f(R_{jl},\theta_{jl})\ ,
\eeq
equation~(\ref{Eq:coherence_i_weakfield}) becomes:
\beq\label{Eq:coherence_i_weakfield2}
\frac{d\sigma_{ge}^j(t)}{dt}= -\left({\Gamma\over 2}+i\Delta\right)\sigma_{ge}^j(t)-\frac{i}{2}\left[\Omega_+(r_j,z_j,t) + \Omega_-(r_j,z_j,t)\right]\exp\left(-ikz_j\right)\ .
\eeq
We performed the calculations presented in this Section keeping both $\Omega_+$ and $\Omega_-$. We have then checked numerically that for our experimental parameters, the field propagating in the direction $-{\bf k}_L$ is negligible, and we will therefore use in the following $\Omega({\bf R}_j,t)\approx\Omega_+({\bf R}_j,t)$.

To proceed, we use a continuous medium approximation, i.e. we calculate the configuration average of both $\sigma_{ge}^j$ and $\Omega_+({\bf R}_j,t)$. In doing so, we explicitly calculate the {\it coherent} part of the electromagnetic field $\langle \Omega_+\rangle$ defined by :
\begin{equation*}
\langle \Omega^*_+({\bf R}_j,t)\rangle = d\langle \left({\bf E}({\bf R}_j,t)\cdot {\bf e}^*_+ \right)\exp\left(-ikz_j\right)\rangle/\hbar\ .
\end{equation*}

This approximation is consistent with the fact that we measure the field in the forward direction, which is dominated by the coherent part in the low-intensity limit. We therefore neglect the fluctuations of the electromagnetic field around its mean. In order to keep the notations simple in what follows,  we will drop the brackets of the configuration average, and call $\Omega_+(r,z,t)$ this {\it coherent} field ($r$ and $z$ are radial and cylindrical coordinates). Similarly, we replace the coherence $\sigma_{ge}^j$ by a continuous function
$\sigma_{ge}(r,z,t)$. Furthermore, we assume that owing to the cylindrical geometry of the cloud and the fact that the transverse dimension $\sigma_r$ is much smaller than the laser waist $w$, the  coherence does not depend on the radial coordinate $r$: $\sigma_{ge}(r,z,t)\approx\sigma_{ge}(z,t)$, making our system effectively one-dimensional.
The central idea of the following derivation is to show that the coherent field $\Omega_+(r,z,t)$ fulfills a paraxial equation, which we will derive. In order to do so, we proceed in several steps. First, we introduce the radial average $\bar{\Omega}_+(z,t)$ of the field $\Omega_+(r,z,t)$, with
\beq\label{Eq:Radial_av_Omega}
\bar{\Omega}_+(z,t)={1\over 2\pi\sigma_r^2}\int_0^\infty\Omega_+(r,z,t) \exp\left[-{r^2\over 2\sigma_r^2}\right]\, 2\pi rdr\ .
\eeq
We then replace the sums by integrals involving the spatial density
distribution of the cloud (with peak density $n$), assumed to be Gaussian
with rms dimensions $\sigma_r$ and $\sigma_z$ in the radial and
longitudinal directions respectively:
\begin{align}
&\bar{\Omega}_+(z,t)={1\over 2\pi\sigma_r^2}\int_0^\infty\Omega_{\rm L}(r,z)\exp\left[-{r^2\over 2\sigma_r^2}\right]\, 2\pi rdr
~+~{n\over 2\pi\sigma_r^2}{3\over4}\Gamma \exp\left(ikz\right)\nonumber\\
& \times \int dx dy \exp\left[-{r^2\over 2\sigma_r^2}\right]\int_{-\infty}^z dz' \int dx' dy'
f(R',\theta') \sigma_{ge}(z',t-\frac{R'}{c})\exp\left(  -ikR'\right)
\exp\left[-{x'^2+y'^2\over 2\sigma_r^2}-{z'^2\over 2\sigma_z^2}\right]\ .
\end{align}
Here, $r=\sqrt{x^2+y^2}$ and $R'=\sqrt{(x-x')^2+(y-y')^2+(z-z')^2}$. Introducing the relative coordinates $X=x-x'$, $Y=y-y'$, and $Z=z-z'$, allows us to re-arrange the integral
and finally, using the expression~(\ref{Eq:laser_field}) for the Gaussian laser field we obtain:
\beq\label{Eq:omegap_continu}
\bar{\Omega}_+(z,t)={z_{\rm R}\over z_{\rm R}-iz +k\sigma_r^2}\Omega_0 +{3\Gamma n\over 8}\exp\left(ikz\right) \int_{-\infty}^z dz' \sigma_{ge}(z',t-\frac{z-z'}{c}) J(z-z')\exp\left[-{z'^2\over 2\sigma_z^2}\right]\ ,
\eeq
with $J(Z)$ a kernel function:
\beq
J(Z)=\int dX dY f(R',\theta') \exp\left[-{X^2+Y^2\over 4 \sigma_r^2}-ikR' \right]\ ,
\eeq
where $R'=\sqrt{X^2+Y^2+Z^2}$.
After a lengthy calculation, we find an  analytical expression for this function:
\beq\label{Eq:Kernel_calculation}
J(z)={J_1(z)\over 2}\left(3+{1\over 2k^2\sigma_r^2}+{k^2z^2\over 4k^4\sigma_r^4}
\right)
+{i\pi\over k^2}\left({1}+i{z\over 2 k\sigma_r^2 }\right)\exp(-ikz)
\eeq
where
\beq
J_1(z)= \int_0^\infty \exp\left(-{r^2\over 4\sigma_r^2}\right)
{\exp(-ik\sqrt{r^2+z^2})\over k\sqrt{r^2+z^2}} \, 2\pi rdr
={2\pi\sqrt{\pi}\over k}\sigma_r\, \exp\left(-k^2\sigma_r^2\right) {\rm Erfc}\left({|z|\over 2\sigma_r}+
i k \sigma_r\right)\exp\left({z^2\over 4\sigma_r^2}\right)\ .
\eeq

Using the asymptotic expression ${\rm Erfc}(x)\approx {1\over x\sqrt{\pi}}\exp(-x^2)$ for $|x|\gg 1$, and considering the fact that $1/kb\leq 0.1$ for our experimental parameters, we get :
\beq\label{Eq:Japp}
J(z)\approx -{2i\pi\over k^2} {b\over b-iz} \exp[-ikz]\ {\rm with }\ b= 2k\sigma_r^2\ .
\eeq

\begin{figure}
\includegraphics[width=0.95\linewidth]{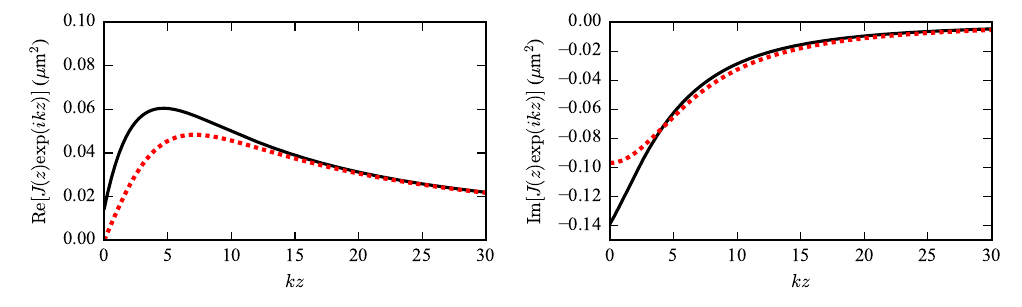}
\caption{(Left) Real part of the function $J(z)\exp(ikz)$ for the exact expression (\ref{Eq:Kernel_calculation}) (solid black line)
and the approximate expression (\ref{Eq:Japp}) (dotted red line). (Right) Same for the imaginary part.}
\label{Fig_SM_Kernel}
\end{figure}

Figure~\ref{Fig_SM_Kernel} compares the exact expression of the kernel $J(z)$ (Eq.\,\ref{Eq:Kernel_calculation}) to the approximate one of Eq.\,(\ref{Eq:Japp}). Both are in excellent agreement for $kz\gtrsim 5$, and deviate at shorter distances.  Hence, replacing in Eq.\,(\ref{Eq:omegap_continu}) $J(z-z')$ by its approximation (\ref{Eq:Japp}) is valid when $k|z-z'|\gtrsim 5$. Since the transverse size of the atomic cloud fulfills $k\sigma_r\gtrsim 1$, this is equivalent to $n/k^3\lesssim 0.2$.
For the experiment presented in the main text $n/k^3\leq 0.15$; we therefore use the approximation (\ref{Eq:Japp}) from now on. Note that this approximation would break down
for denser clouds.

The function ${b\over b-iz}$ can be written as a radial average:
\begin{align}
{b\over b-iz}={1\over 2\pi \sigma_r^2}\int_0^\infty \,
\exp\left[-{r^2\over 2\sigma_r^2}\right]{2b\over b-2iz}\exp\left[-\frac{kr^2}{b-2iz}\right]\, 2\pi r dr\ .
\end{align}
As a consequence, the above calculation shows that the field $\bar{\Omega}_+(z,t)$ is the radial average of the field
\beq
\Omega_+(r,z,t)=\Omega_{\rm L}(r,z) -i{3\pi \Gamma n\over 4 k^2} \int_{-\infty}^{z}dz'\, \sigma_{ge}(z',t-{z-z'\over c}){2b\over b-2i(z-z')}
\exp\left[-{kr^2\over  b-2i(z-z')}\right]\exp\left[-{z'^2\over 2\sigma_z^2}\right]\exp\left(ikz'\right) \ .
\eeq
In this form, one can now check that the coherent field $\Omega_+(r,z,t)$ is solution of a paraxial equation, as advertised above:
\beq\label{Eq:Paraxial_eq}
{\partial \Omega_+(r,z,t)\over \partial z}+{1\over c}{\partial \Omega_+(r,z,t)\over \partial t}-{1\over 2ikr}{\partial\over \partial r}\left[r{\partial\Omega_+(r,z,t)\over \partial r}\right]
= -i{3\pi\over2 k^2}\Gamma n\, \sigma_{ge}(z,t)
\exp\left[- {kr^2\over b}-{z^2\over 2\sigma_z^2}\right]\exp(ikz)\ .
\eeq
This equation is generally derived in the plane-wave approximation,
which corresponds to large values of the parameter $b$ and $z\ll b$ (see~\cite{Feld1971} and references therein).
Here, we have demonstrated it for an ensemble of two-level atoms, but it could be extended to degenerate two-level
systems~\cite{Crubellier1986} and to more complex configurations~\cite{RubinszteinDunlop2001,Agarwal2000}.
Importantly, Eq.\,(\ref{Eq:Paraxial_eq}) would still be valid in the strong driving regime.
Note also that to reach this equation, it was necessary to keep the retarded times $t-R/c$
in the Eq.(\ref{Eq:Born_without_Markov}): extra terms would have appeared otherwise.
%, {\it i.e.} not  performing the Markov approximation.
%Equation (\ref{Eq:Paraxial_eq}) would not look as simple had we done the usual Markov approximation.
This paraxial equation for the field, is coupled to the equation on the coherence:
\beq\label{Eq:coherence_continu}
{\partial \sigma_{ge}(z,t)\over \partial t}= -\left({\Gamma \over 2}
+i\Delta\right)\sigma_{ge}(z,t) - i\frac{\Omega_+(r,z,t)}{2}\exp\left(-ikz\right)\ ,
\eeq
with the initial conditions $\Omega_+(r,z=-\infty,t)=\Omega_{\rm L}(r,z=-\infty,t)$ and $\sigma_{ge}(z,t=0)=0$. However, owing to the radial dependence $r$, the  equations cannot
be easily solved.

We then consider the diffraction of the light by the cloud.
Diffraction transfers part of the laser light
with gaussian spatial profile to higher order Gaussian modes.
Mathematically, we thus decompose the field
$\Omega_+(r,z,t)$ onto the Laguerre-Gauss
modes propagating in the direction ${\bf k}_{\rm L}$:
\beq\label{Eq:decomp_LG}
\Omega_+(r,z,t)= \sqrt{\pi z_{\rm R}\over k}\sum_{q=0}^{\infty} \Omega_+^{(q)}(z,t){\rm LG}_q(r,z)\ .
\eeq
The expressions of the Laguerre-Gauss modes are:
\beq
{\rm LG}_q(r,z)=\sqrt{k\over \pi z_{\rm R}}\left(\frac{1+iz/z_{\rm R}}{1-iz/z_{\rm R}}\right)^q\left({1\over 1-iz/z_{\rm R}}\right){\rm L}_q\left( {kz_{\rm R}r^2\over z^2+z_{\rm R}^2}\right)\exp\left(-\frac{i}{2}{kr^2\over (z+iz_{\rm R})}\right)\ ,
\eeq
with ${\rm L}_q(r)$ the Laguerre polynomial of order $q$. The mode $q=0$ describes the Gaussian mode of the probe field.
The cylindrical symmetry of our problem allows us to restrict to the modes
with $m=0$ only. They follow the orthogonality conditions:
\beq\label{Eq:orthoLG}
\int_0^\infty {\rm LG}_q(r,z)  {\rm LG}_p^*(r,z)2\pi rdr=\delta_{p,q}\ .
\eeq

As a last step, we perform a radial average similar to the one of Eq.\,(\ref{Eq:Radial_av_Omega}).
Defining the function $f_q(z)$ as~\cite{Footnote_fq}:
\beq
f_q(z)= \int_0^\infty \exp\left( - {r^2\over 2\sigma_r^2}\right)
{\rm LG}_q(r,z)2\pi r dr\ ,
\eeq
and using the fact that the functions ${\rm LG}_q(r,z)$ are solutions of a paraxial equation, we reach a set of coupled equations that are now easy to solve numerically:
\beq\label{Eq:MB_coherence}
{\partial \tilde{\sigma}_{ge}\over \partial t}(z,t)= -\left({\Gamma\over 2}+i\Delta \right)\tilde{\sigma}_{ge}(z,t) -i\frac{\underline{\Omega}(z,t)}{2}
\eeq
and
\beq\label{Eq:MB_Omega}
{\partial \Omega_+^{(q)}\over \partial z} + {1\over c}{\partial \Omega_+^{(q)}\over \partial t}=
-i\sqrt{2\over \pi w^2}{3\pi\over 2 k^2}\Gamma n  f^*_q(z) \exp\left(-{z^2\over 2\sigma_z^2}\right)\, \tilde{\sigma}_{ge}(z,t)\ ,
\eeq
with $\tilde{\sigma}_{ge}(z,t)=\sigma_{ge}(z,t)\exp(ikz)$, the slowly-varying coherence and $\underline{\Omega}(z,t)$, the Rabi frequency given by:
\beq\label{Eq:UnderlinedOmega}
\underline{\Omega}(z,t)={\sqrt{\frac{\pi w^2}{2}} {1 \over 2\pi \sigma_r^2}}\sum_{p=0}^{\infty} f_p(z)\Omega_+^{(p)}(z,t)\ ,
\eeq

We make the connection with the experimentally measured quantity by noticing that the fibered avalanche photo-diode measures the projection of the total field $\Omega_+(r,z,t)$ at the position of the lens onto the Gaussian mode of the single-mode fiber. The avalanche photo-diode therefore measures $|\Omega_+^{(0)}(z,t)|^2$, with $z$ a distance that we take in practice equal to $10\sigma_z$.

Finally, the steady-state response is solution of the coupled differential equations:
\begin{align}\label{Eq:MB_steady}
{d\Omega_+^{(q)}\over dz}(z)=-{3\Gamma n\over 4(k \sigma_r)^2}{f^*_q(z)\over \Gamma + 2i\Delta}\,\exp\left(-{z^2\over 2\sigma_z^2}\right)\sum_{p=0}^{\infty}f_p(z)\Omega_+^{(p)}(z)\ .
\end{align}
Equation~(\ref{Eq:MB_steady}) is therefore a continuous version of the coupled-dipole equations used to calculate the optical response of the cloud. However, the number of equations
to solve does not depend on the number of atoms $N$, but only on the number of Laguerre-Gauss modes involved. In practice we have found that 5-10 modes are enough. Therefore, beyond providing a framework, which can be extended to the case of high intensity of the driving field, our model is less demanding computationally than microscopic models, such as the coupled-dipole model.

As a final note, if the longitudinal position of the cloud is displaced by $z_0$ with respect to the position of the beam waist at $z=0$, the only modification consists in replacing $z$
by $z-z_0$ only in the $\exp\left(-{z^2\over 2\sigma_z^2}\right)$ term in Eq.\,(\ref{Eq:MB_steady}).

\section{Derivation of the multi-mode Maxwell-Bloch model: strong intensity case.}\label{Sec:details_MB_strong_field}

In this Section, we give without derivation the main equations of the model in the strong driving limit. Although they are
valid for any intensity of the driving laser, the price to pay is to ignore the correlations between atoms.
%, {\it i.e.} resort to
%a mean-field approximation.

The set of equations governing the coherence $\sigma_{ge}(z,t)$, the  population  $\sigma_{ee}(z,t)$ and the component of the field $\Omega_+^{(q)}$ is
(the equation for the field is identical to the one in the low intensity case):
\begin{eqnarray}\label{Eq:MB_strong}
{\partial \sigma_{ge}\over \partial t}(z,t)&=&-\left({\Gamma\over 2}+i\Delta \right)\sigma_{ge}(z,t)-i\frac{\underline{\Omega}(z,t)}{2}\exp(-ikz)[1-2\sigma_{ee}(z,t)]\ , \\
{\partial \sigma_{ee}\over \partial t}(z,t)&=&
-\Gamma \sigma_{ee}(z,t)+\frac{i}{2}\left[\underline{\Omega}^*(z,t)\exp(ikz)\sigma_{ge}(z,t)-\underline{\Omega}(z,t)\exp(-ikz)\sigma_{eg}(z,t)\right]\ , \\
{\partial \Omega_+^{(q)}\over \partial z} + {1\over c}{\partial \Omega_+^{(q)}\over \partial t}&=& -i\sqrt{2\over \pi w^2}{3\pi\over 2k^2}\Gamma n  f^*_q(z)
\exp\left(-{z^2\over 2\sigma_z^2}\right)\, \sigma_{ge}(z,t)\exp(ikz)\ ,
\end{eqnarray}
with the Rabi frequency $\underline{\Omega}(z,t)$ given by Eq.(\ref{Eq:UnderlinedOmega}).
%\beq
%\underline{\Omega}(z,t)=\sqrt{\pi w^2\over 2}{1 \over 2\pi \sigma_r^2}\sum_{p=0}^{\infty}f_p(z)\Omega_+^{(p)}(z,t)\ .
%\eeq

To obtain these equations, we used the same approximations as for the low field limit:
(i) neglect the counter-propagating field $\Omega_-$, and
(ii) use the one-dimensional approximation: $\sigma_{ge}(r,z,t)\approx \sigma_{ge}(z,t)$ and
$\sigma_{ee}(r,z,t)\approx \sigma_{ee}(z,t)$.
Importantly, when we performed the continuous approximation,
starting from Eqs.~(\ref{Eq:coherence_i_Appendix}) and (\ref{Eq:population_i_Appendix}),
we had to %rely on a mean-field approximation where we
neglect the correlations between the coherences
and the populations of different atoms:
$\langle\sigma_{ge}^l\sigma_{ge}^j\rangle\approx\langle\sigma_{ge}^l\rangle\langle\sigma_{ge}^j\rangle$, and
$\langle\sigma_{ge}^l\sigma_{ee}^j\rangle\approx\langle\sigma_{ge}^l\rangle\langle\sigma_{ee}^j\rangle$,
with $\langle\cdot \rangle$ the configuration average resulting from the continuous approximation.
We note that such an  approximation {\it is not} necessary
in the low intensity, {\it i.e.} classical, regime. Finally, as in the low-intensity regime,
the continuous approximation
explicitly neglects the fluctuations of the field.

The steady-state regime in the strong intensity case is then governed by the following equation:
\beq\label{Eq:MB_steadystate_strong}
{d\Omega_+^{(q)}\over dz}(z)=-{3\Gamma n\over 4(k \sigma_r)^2}{f^*_q(z)\over \Gamma + 2i\Delta}\,\exp\left(-{z^2\over 2\sigma_z^2}\right)
\sum_{p=0}^{\infty} f_p(z)\Omega_+^{(p)}(z)\, {\Gamma^2+4\Delta^2\over \Gamma^2+4\Delta^2+ 2|\underline{\Omega}(z,t)|^2}\ .
\eeq
When the driving intensity is small ($\Omega_0\ll \Gamma$ hence $|\underline{\Omega}(z,t)|\ll \Gamma$), one recovers Eq.\,(\ref{Eq:MB_steady}).

\section{Coupled dipole model}\label{Appendix:detailed_coupled_dipoles}

In this last Section, we use classical electrodynamics to derive
the  coupled-dipole equations  for our experimental conditions
of an ensemble of $\sigma_+$-polarized, two-level atoms driven by a laser field with
a linear polarization perpendicular to the quantization axis
(see the geometry in Fig. 1(a) of the main text).

We first assign a polarizability $\alpha(\omega)$ to each two-level atom, given in the near-resonant approximation by:
\beq
\alpha(\omega)={6\pi i\over k^3}{1\over 1-i{2 \Delta\over \Gamma}}\ ,
\eeq
with $\Delta=\omega-\omega_0$ the detuning between the laser frequency $\omega$ and the atomic transition $\omega_0$ between states $|g\rangle$ and $|e\rangle$.
The $\sigma_+$-polarized dipole ${\bf d}_j$ of atoms $j$ is given by ${\bf d}_j=d_j {\bf e}_+$, with
${\bf e}_+=-({\bf e}_y+i{\bf e}_z)/\sqrt{2}$
(${\bf e}_x$ being the unit vector along the quantization axis set by the magnetic field). The field scattered by the $\sigma_+$-polarized dipole of atom $l$
onto the $\sigma_+$-polarized dipole $j$, both separated by a distance $R_{jl}$, is~\cite{Jackson}:
\beq\label{Eq:dipdip}
{\bf E}_{l\rightarrow j}=
{k^3 d_l\over 4\pi\epsilon_0}{e^{i kR_{jl}}\over (k R_{jl})^3}
\left[ (3(\hat{\bf r}_{jl}\cdot {\bf e}_+)\hat{\bf r}_{jl}-{\bf e}_+)(1-i k R_{jl})
+ (\hat{\bf r}_{jl}\times {\bf e}_+)\times\hat{\bf r}_{jl}(kR_{jl})^2\right] \ ,
\eeq
with $\hat{\bf r}_{jl}={\bf R}_{jl}/R_{jl}$.
In steady-state, the dipole ${\bf d}_j$ of dipole $j$ is driven by the laser field ${\bf E}_{\rm L}({\bf R}_j)=E_{\rm L}({\bf R}_j) {\bf e}_y$ and the field scattered by all the other atoms:
\beq
{\bf d}_j=\epsilon_0\alpha(\omega) [{\bf E}_{\rm L}+
\sum_{l\neq j}{\bf E}_{l\rightarrow j}]
\eeq
The complex amplitudes $d_j={\bf d}_j\cdot {\bf e}_+^*$
of the dipoles are therefore solution
of  the set of coupled equations:
\beq\label{Eq:coupled_dipole}
(i\Delta-{\Gamma\over 2}) d_j =
\epsilon_0 {3i\pi \Gamma\over k^3}{E_{\rm L}({\bf R}_j)\over \sqrt{2}}+
i\sum_{l=1, l\neq j}^N {V_{jl}\over \hbar} d_l\ ,
\eeq
with the dipole-dipole interaction term given by:
\beq\label{Eq:dipdip}
V_{jl}=-{3\over 8}{\hbar\Gamma\over (k R_{jl})^3}e^{i kR_{jl}}
\left[ (1-3\cos^2\theta_{jl})(1-i k R_{jl} ) + (1+\cos^2\theta_{jl})(kR_{jl})^2\right] \ ,
\eeq
with $\theta_{jl}$ the angle between ${\bf R}_{jl}$ and the quantization axis ${\bf e}_x$.
Equation~(\ref{Eq:coupled_dipole}) is equivalent to Eq.\,(\ref{Eq:coherence_i_weakfield})
in steady-state, using the relation between the complex
amplitude of the dipole and the optical coherence: $d_j=2d\, \sigma_{eg}^j$.

To model the data, we  use a stochastic approach analog to the one we used in our previous works~\cite{Pellegrino2014,Jennewein2016a,Jenkins2016b}.
We first solve the set of coupled dipole
equations for a given realization of the
spatial distribution of atoms and calculate the scattered electric field,
in the far-field, using:
\beq
{\bf E}_{\rm sc}={k^2\over 4\pi \epsilon_0}\sum_{j=1}^N [(\hat{\bf r}_j\times
{\bf e}_+)\times\hat{\bf r}_j]{d_j\over R_j}e^{i kR_j}\ ,
\eeq
with $\hat{\bf r}_j={\bf R}_j/R_j$, ${\bf R}_j$ being the distance between the atom $j$
and the position of observation in the plane of the lens.
We compute the total field ${\bf E}= {\bf E}_{\rm L}+ {\bf E}_{\rm sc}$,
using the far-field expression of the laser field
\beq
{\bf E}_{\rm L}= -i{z_{\rm R}\over R} E_0 e^{ikR} {\bf e}_y\ .
\eeq
We then form the quantity ${\bf E}\cdot{\bf E}_{\rm L}^*$
and integrate over the surface of the lens to get the
transfer function $S(\omega)$:
\beq\label{Eq:TransferS}
S(\omega) = {\int {\bf E}({\bf r},\omega)\cdot {\bf E}_{\rm L}^*({\bf r})dA
\over  \int  |{\bf E}_{\rm L}({\bf r})|^2 dA}\ .
\eeq
We finally average the transfer function over 100 realizations
of the spatial distribution of the atoms in the cloud, to get the coherent transfer function.